# Analysis of fluid flow in fractal microfluidic channels

Jun Hu[1,2], Zhan-Long Wang[1]*

[1]Shenzhen Institute of Advanced Technology, Chinese Academy of Sciences, Shenzhen, Guangdong 518000, China.

[2]China University of Political Science and Law, Beijing, 100091, China.

*Corresponding author: Zhan-Long Wang (zl.wang1@siat.ac.cn)

**Abstract**

Fractal channels have significant applications in fields such as microfluidic chips and in vitro diagnostics. However, there is currently insufficient understanding and recognition of fluid flow within fractal channels. In this paper, the fluid flow within dendritic fractal structures was studied with finite element analysis (FEA), focusing on fluid velocity and mass fraction variations. The research introduces the ratio RA (the ratio of the longitudinal channel length to the lateral channel length) as a key parameter to evaluate fractal structures. A smaller RA corresponds to a flatter fractal configuration. By comparing flow rate, velocity, and mass distribution at the outlet of fractal channels with varying RA values, the results demonstrate that flatter fractal structures lead to more uniform flow distribution at the outlet. Meanwhile, the uniformity of fluid flow at the outlet of the fractal channels was analyzed. This work indicates that channel geometry significantly influences fluid dynamics, and also provides valuable insights into optimizing flow dynamics in small-scale applications, contributing to the design of more efficient fluidic devices.

## 1. Introduction

Microfluidic technology has transformed a wide range of scientific and industrial fields, providing critical advancements in microfluidic chips, organ-on-a-chip systems, and in vitro diagnostics. By miniaturizing fluidic processes, microfluidic systems enable precise control over the manipulation of small volumes of fluids, leading to innovations in areas such as chemical synthesis, biological analysis, and medical diagnostics[1,2]. In particular, microfluidic chips, also known as lab-on-a-chip (LOC) devices, have become indispensable tools in biomedical research. These devices integrate complex laboratory functions onto a small platform, enabling rapid testing, reduced reagent use, and enhanced sensitivity in applications like cancer diagnostics and cell-based assays[3,4]. LOC devices have been utilized for detecting cancer-derived biomarkers, such as extracellular vesicles (EVs) and circulating tumor cells (CTCs), as well as for performing drug screenings and personalized medicine assessments[5]. Beyond diagnostics, microfluidic platforms play a critical role in the development of organ-on-a-chip technologies, which mimic the functions of human organs in a controlled environment[6]. These devices have been used to model the heart, lung, liver, and other organs, offering potential applications in drug toxicity testing, disease modeling, and tissue engineering[7,8]. Another key area where microfluidics has made strides is in vitro diagnostics, where it offers highly sensitive detection systems for pathogens and biomolecules[9,10]. Microfluidic chips are becoming central to rapid diagnostic platforms used in clinical settings, particularly for detecting viruses, bacteria, and other infectious agents[11]. The enhanced capabilities of microfluidic systems, including high-throughput testing, point-of-care diagnostics, and integration of biosensors, demonstrate their broad potential across many domains[12]. As the demand for more complex and efficient microfluidic systems increases, new designs and structures are being explored to optimize their performance, with fractal microchannel structures emerging as one of the most promising innovations. Among the numerous designs and structures employed in microfluidics, dendritic or fractal microchannel structures have garnered significant attention due to their unique properties and versatile applications. These fractal structures, characterized by their self-similar and scale-invariant features, offer enhanced performance in mixing, reaction, heat transfer, and other critical processes in microfluidic systems[13].

In recent years, dendritic fractal microchannel structures have gained significant attention for their unique ability to enhance fluid dynamics in microfluidic applications. Characterized by their self-similar and scale-invariant properties, fractal microchannels offer improved mixing, reaction, and heat transfer processes by creating complex flow paths that maximize surface area[14,15]. This has led to their increasing incorporation into microfluidic chips, with applications in various fields such as cancer diagnostics, cell-based assays, and biomolecule detection[16-19]. For example, one notable application of dendritic fractal microchannel structures is in the enhancement of cancer diagnostics. By employing fractal designs, microfluidic chips can achieve better separation and enrichment of circulating tumor cells (CTCs), extracellular vesicles (EVs), and circulating DNA (ctDNA), significantly improving the efficiency and sensitivity of liquid biopsy techniques[20,21]. Fractal microchannels also enhance the capture and analysis of cancer biomarkers, allowing for more precise and quicker diagnostics compared to conventional methods[22]. Similarly, dendritic fractal designs are being employed in cell-based assays to study immune cell interactions, particularly in the context of anti-tumor responses[23]. These structures offer high-throughput capabilities for detailed analyses of cell-cell interactions, aiding in the understanding of immune system dynamics and their therapeutic potential[24]. In addition to cancer diagnostics, fractal microchannel structures have been applied to in vitro diagnostic systems aimed at the detection of a wide range of biomolecules. For example, fractal silver dendrites have been used to create 3D surface-enhanced Raman scattering (SERS) platforms, which demonstrate exceptional sensitivity in detecting biomolecules under hydration conditions[25]. These platforms provide enhanced optical properties due to the large surface area and resonant structures inherent to fractal designs, enabling the detection of low concentrations of target molecules in biomedical research. Similarly, fractal microchannels have been integrated into immunofluorescence-based assays for virus detection, such as microfluidic dielectrophoresis (DEP) chips for rapid dengue virus detection[26,27]. These platforms offer higher capture efficiency and faster immunoreactions, which are crucial for timely clinical diagnostics. Moreover, fractal microchannel designs have found applications beyond diagnostics, particularly in drug discovery and material synthesis. By enhancing fluid mixing and reaction control, fractal microchannels enable more efficient chemical processes in microfluidic devices[28]. The miniaturization of these systems allows for high-throughput screening and synthesis in pharmaceutical research, offering advantages over traditional macroscopic systems[29]. As such, fractal microchannel structures continue to show promise in a wide range of fields, where they enhance the performance of microfluidic systems in ways that conventional designs cannot match.

Despite the substantial advancements in the application of dendritic fractal microchannel structures in microfluidics, a comprehensive understanding of fluid flow within these structures remains incomplete[30]. Many studies have demonstrated the benefits of fractal designs in enhancing mixing, separation, and reaction efficiency, but there is still a need for more detailed investigations into the behavior of fluids within these complex geometries[31]. In the past, research has shown that the hydrodynamic performance of these networks, composed of series of rough ducts, varies significantly between laminar and turbulent flow regimes. The transient response of internal fluid pressure within these structures exhibits oscillatory behavior, which is closely linked to the characteristics of both the fluid and the network itself[32]. Additionally, the morphology of natural micro-fractures in coal and their impact on fluid flow have been investigated using fractal theory and the lattice Boltzmann method. The results indicate that the presence of dominant channels significantly enhances permeability, while orthogonal micro-fracture networks are less conducive to fluid flow[33]. The behavior of fluid flow during imbibition in regular porous media has also been studied, revealing that the invasion morphology can transition from compact and faceted to irregular and dendritic, depending on the porosity and wettability of the media. This transition is reflected in the fractal dimension of the invasion patterns, providing new insights into the role of geometrical features in multiphase flow[34]. In studies related to interface wetting, phenomena such as droplet spreading, spontaneous droplet motion, and contact line pinning have been widely observed and analyzed[35-41]. These mechanisms are also significant in interface condensation processes, where the wetting properties of a surface can drastically influence fluid flow behavior. Understanding these effects is crucial for designing structures that optimize fluid distribution and control, such as in fractal channels and microfluidic devices. Monte-Carlo simulations have further elucidated the influence of fluid flow on the morphological development of solidification structures. Laminar flow tends to destabilize the solid-liquid interface, promoting dendritic growth, while turbulent flow can produce fine and compact structures[42]. Understanding the flow dynamics, including velocity profiles, mass fraction distribution, and outlet flow rates, is critical for optimizing the design and application of fractal microchannels in practical systems. Additionally, while fractal structures have been widely adopted for their advantages in fluid manipulation, the impact of varying structural proportions—particularly the ratio of longitudinal to lateral channel lengths—on fluid behavior has not been thoroughly explored.

In this study, we aim to fill this knowledge gap by conducting a detailed analysis of fluid flow within dendritic fractal microchannels using finite element analysis (FEA). Specifically, we introduce the ratio RA, which represents the ratio of the longitudinal channel length to the lateral channel length, as a key parameter for evaluating fractal structures. A smaller RA value corresponds to a flatter configuration, which we hypothesize will lead to more uniform flow distribution at the channel outlet. By varying RA and analyzing the resulting flow dynamics, including velocity, mass fraction, and flow rate, we provide a comprehensive assessment of how fractal channel geometry influences fluid behavior. The insights gained from this study will not only contribute to a deeper understanding of fractal microchannel flow but also inform the design of more efficient microfluidic devices for applications in diagnostics, drug development, and other fields where precise fluid control is essential.

## 2. Materials and methods

In this study, we employed a conventional PDMS microfluidic chip structure to simulate fluid flow within a dendritic fractal network using the finite element method.

The theoretical framework for our simulations is based on the Navier-Stokes (NS) equations, which govern the motion of fluid under the influence of viscous and inertial forces. The NS equations are commonly used to describe fluid flow behavior and are particularly well-suited for the study of microfluidic systems, where precise control over flow parameters is essential for various biological and chemical applications. To simulate the interface dynamics between the fluid and the microchannel walls, the level set method was employed. This method is well-established for accurately capturing the fluid interface, especially in cases where complex geometries like dendritic fractal networks are involved. The conservation of mass equation (continuity equation) is given by:

$$\nabla \cdot \boldsymbol{u} = 0,$$

where $\boldsymbol{u}$ represents the velocity vector field, ensuring mass conservation throughout the flow. The NS equation (Conservation of momentum) is given by:

$$\rho\left(\frac{\partial \boldsymbol{u}}{\partial t} + (\boldsymbol{u} \cdot \nabla)\boldsymbol{u}\right) = -\nabla p + \mu \nabla^2 \boldsymbol{u} + \boldsymbol{f},$$

where $\rho$ is the fluid density, $\mu$ is the dynamic viscosity, $p$ is the pressure, and $\boldsymbol{f}$ represents external body forces acting on the fluid. These equations govern the fluid motion, capturing the effects of viscosity, inertia, and pressure gradients, and are essential for characterizing flow patterns within the fractal microchannels.

To accurately capture the fluid interfaces and track the movement of fluid boundaries within the dendritic fractal structure, the level set method is employed. The level set method is a powerful tool for modeling moving interfaces, such as the boundary between two immiscible fluids or the fluid-air interface. The interface is represented as the zero level set of a scalar function $\phi$, which evolves over time according to the level set equation:

$$\frac{\partial \phi}{\partial t} + \boldsymbol{u} \cdot \nabla \phi = 0,$$

where, $\phi$ represents the signed distance from the interface, with positive values indicating one fluid phase and negative values indicating the other. This approach allows for the efficient tracking of complex interface geometries and ensures smooth and accurate interface evolution during the simulation.

The microfluidic chip used in this study is made of PDMS with a contact angle of 120 degrees, which implies that the surface is hydrophobic. The solution inside the microchannels contains doxorubicin, a widely used chemotherapeutic agent, ensuring relevance to biomedical applications. At the inlet, a fixed flow rate was applied to control the flow, while the fluid velocity was initially set to zero to simulate a no-flow starting condition. The boundary conditions applied to the system include a no-slip condition at the channel walls, meaning the velocity of the fluid adjacent to the walls is zero. For the outlet, we applied a fixed pressure condition to simulate open flow exiting the microchannel. The boundary conditions ensure that the simulation mimics real-world microfluidic behavior, particularly in drug delivery and diagnostic systems where controlled flow is critical. The simulation was performed using the finite element method (FEM), which discretizes the domain into small elements, allowing for the numerical solution of the governing equations. The fractal microchannel was modeled as a 2D structure, with the flow properties, such as velocity distribution, mass fraction, and outlet flow rates, calculated over time. The diagram of the structure is depicted in Fig. 1. The fluid inlet and outlet were illustrated in Fig. 1a. A simulation diagram was displayed in Fig. 1b. By varying the aspect ratio of the fractal channels, we were able to investigate the impact of structural variations on the uniformity and distribution of fluid flow.

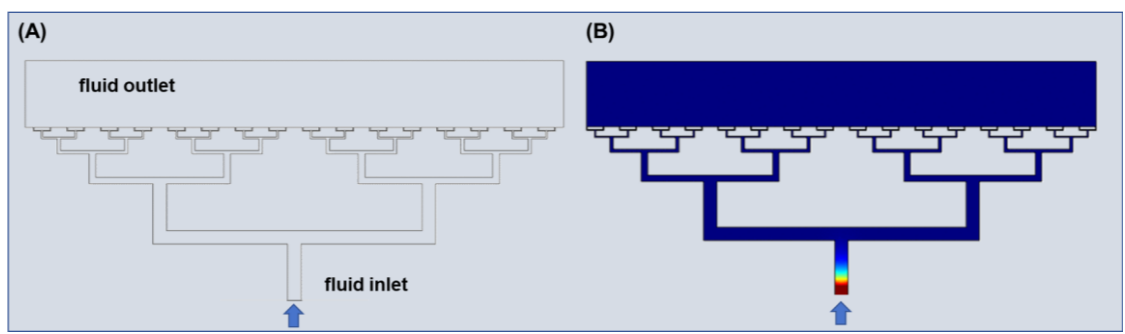

**Figure 1. Schematic representation of the fractal channel structure.** (A) Diagram of the multi-level fractal network with fluid inlet and multiple fluid outlets. The hierarchical structure is designed to ensure efficient

distribution of fluid flow across the outlets. (B) The diagram of finite element simulation of the fluid velocity distribution within the fractal structure.

## 3. Results and discussion

### 3.1. The development and variation of fluid velocity in fractal channels

The fluid velocity distribution within the dendritic fractal network over the course of 0.2 s is depicted in Fig. 2. At time $t = 0$ s (Fig. 2a), the fluid velocity within the entire fractal microchannel structure is zero, as the simulation begins with a no-flow condition. At $t = 0.02$ s (Fig. 2b), a small but noticeable velocity begins to develop within the primary branches of the fractal structure. However, the velocity remains low, indicating that fluid movement is limited to the initial, larger-scale structures of the network at this early stage. As time progresses, fluid motion becomes more pronounced, with slight velocity increases at $t = 0.06$ s (Fig. 2c) and $t = 0.1$ s (Fig. 2d). However, the velocity across the entire network remains relatively low during these stages, suggesting that the fluid has yet to propagate fully through the multi-level fractal channels. By $t = 0.11$ s (Fig. 2e), a significant change occurs, with more noticeable velocity variations, particularly in the larger, primary channels. This marks the point at which the fluid begins to fill the fractal structure more completely, with greater flow activity across multiple levels. At $t = 0.12$ s (Fig. 2f), the fluid velocity has reached the smallest, fifth-order branches of the fractal network, indicating that fluid flow has propagated fully throughout the entire structure. This transition also reflects an observable change at the outlet, where the velocity begins to register meaningful values. From this point onward, in the successive frames of $t = 0.13$ s (Fig. 2g), $t = 0.14$ s (Fig. 2h), $t = 0.15$ s (Fig. 2i), $t = 0.16$ s (Fig. 2j), $t = 0.18$ s (Fig. 2k), and $t = 0.2$ s (Fig. 2l), the velocity distribution across the network stabilizes, with the flow becoming steady. The results clearly show that the highest velocity occurs within the primary, first-order branches of the fractal structure. As the fluid flows into the smaller branches, from the first to the fifth level, the velocity decreases progressively. The velocity in the fifth-order channels is approximately half of that in the first-order branches, illustrating the natural flow attenuation in smaller, more confined sections of the fractal network.

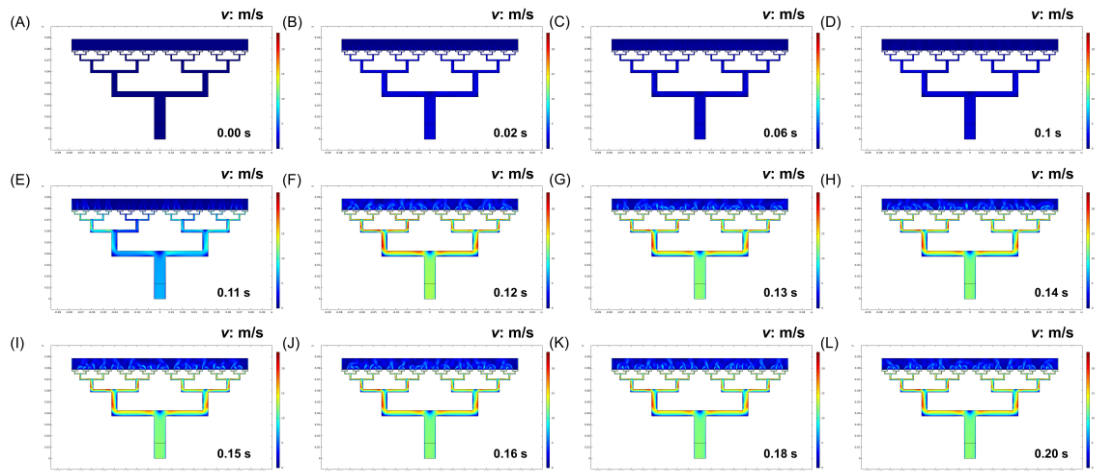

**Figure 2. Fluid velocity distribution in the dendritic fractal network over time.** The sequence of images from $t = 0$ s to 0.2 s shows the progressive development of fluid velocity within the fractal microchannel structure. (A) At $t = 0$ s, no fluid movement is observed, as the initial velocity is set to zero. (B) At $t = 0.02$ s, small velocities are seen in the first-order branches of the fractal network, indicating the onset of fluid motion. (C) By $t = 0.06$ s, fluid velocity increases slightly, but the overall movement remains localized to the larger channels. (D) At $t = 0.1$ s, fluid flow is still relatively slow, with minimal velocity changes observed across the structure. (E) At $t = 0.11$ s, a marked increase in fluid velocity is visible, particularly in the primary channels. This is the point at which the fluid begins to flow more fully into the fractal network. (F) By $t = 0.12$ s, the fluid reaches the fifth-order branches, demonstrating full propagation through the network. (G) and (H) show the steady-state velocity development at t = 0.13 s and $t = 0.14$ s, respectively. The flow stabilizes further at $t = 0.15$ s (I), $t = 0.16$ s (J), $t = 0.18$ s (K), and finally at $t = 0.2$ s (L). In all images, the highest velocity occurs in the first-order branches, with velocities progressively decreasing as the fluid moves into the smaller, fifth-order branches. The steady-state condition suggests efficient fluid distribution throughout the fractal network.

The observed development and evolution of fluid velocity in the dendritic fractal network highlight several key features of fluid dynamics within multi-level channel structures. Initially, the fluid requires time to propagate through the various levels of the fractal structure, as evidenced by the delayed velocity development in the smaller branches. This behavior is typical of hierarchical microchannel systems, where larger channels serve as primary conduits, distributing the fluid into progressively smaller branches. One of the most significant findings from these results is the clear relationship between channel size and fluid velocity. As expected, the larger, primary channels exhibit higher velocities due to their lower flow resistance, while the smaller channels experience reduced velocities as a result of increased viscous effects and the confined nature of the microchannels. This flow attenuation, observed as the fluid moves from the first- to the fifth-order branches, suggests that dendritic fractal networks provide a natural mechanism for distributing fluid uniformly across varying scales. The results also indicate that the fractal design enables the efficient filling of the microchannels, with the flow reaching all levels of the structure within a relatively short time. The steady-state flow observed after 0.12 seconds suggests that the fractal network can achieve stable fluid distribution across all levels, which is essential for applications requiring uniform solution dispersion, such as in drug delivery systems or immunoassays. The reduction in velocity as the fluid reaches smaller branches could also play a role in enhancing reaction times or providing more controlled environments in biological or chemical assays.

**3.2. The flow velocity and mass fraction of the cross-section of outlet**

In this section, we analyze the fluid velocity distribution across the outlet of the fifth-level fractal structure over a period of 1.1 seconds. The study focuses on the velocity variation at a cross-sectional plane taken at the outlet, as depicted in Fig. 3, which highlights the location of the section where velocity measurements were recorded. The outlet of the fractal structure consists of 32 small branches, corresponding to the final fifth-level bifurcations of the fractal network. This complex network arrangement influences the fluid's velocity profile as it exits through multiple microchannels. In Fig. 4, the velocity profiles are shown at different time points from 0 to 1.1 seconds, reflecting how the fluid flow develops and stabilizes across the outlet section. At 0 seconds (Fig. 4a), there is minimal movement within the system, with nearly zero velocity recorded across the outlet, indicating that the fluid has not yet begun to flow significantly into the fractal channels. At 0.1 seconds (Fig. 4b), the fluid velocity begins to rise across the outlet, reaching approximately 0.1 m/s.

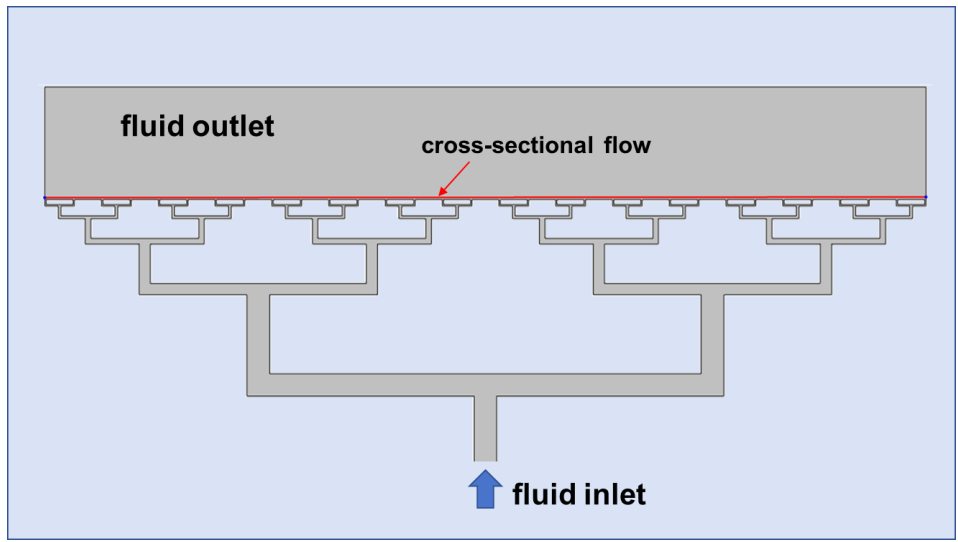

**Figure 3. A schematic representation of the cross-section taken at the outlet of the fifth-level fractal structure.** The section is located at the exit of the fractal network, where fluid flows out from 32 small branching microchannels that represent the fifth-level bifurcations. The red line in the figure indicates the location where the velocity profile was analyzed in subsequent figures.

As time progresses, from 0.2 seconds (Fig. 4c) to 0.4 seconds (Fig. 4e), the velocity increases slightly, reaching a range close to 0.15 m/s. However, the velocity distribution remains relatively uniform, with no significant peaks or variations across the section. The most notable increase in velocity occurs at 0.5 seconds (Fig. 4f), where the velocity starts to exceed 0.2 m/s. From this point onwards, the velocity remains in a stable range between 0.2 m/s and 0.3 m/s, with some variation near the edges of the outlet channels. In particular, by 0.6 seconds (Fig. 4g) and continuing through 0.7 seconds (Fig. 4h) to 1.1 seconds (Fig. 4l), the velocity reaches its peak in certain sections, although the overall distribution remains relatively homogeneous, with higher velocities near the boundary of the outlet branches. The uniformity observed at the early time points suggests a gradual and controlled fluid movement through the fractal structure, where velocity development is relatively steady. The presence of higher velocity near the outlet edges from 0.5 seconds onwards indicates that the geometry of the fractal network influences localized flow behavior, possibly due to the confined nature of the microchannels. This trend also suggests that as the flow continues, there may be a steady-state regime where the velocity remains balanced across the microchannels, with edge effects becoming more pronounced due to the interaction between the fluid and the microchannel walls.

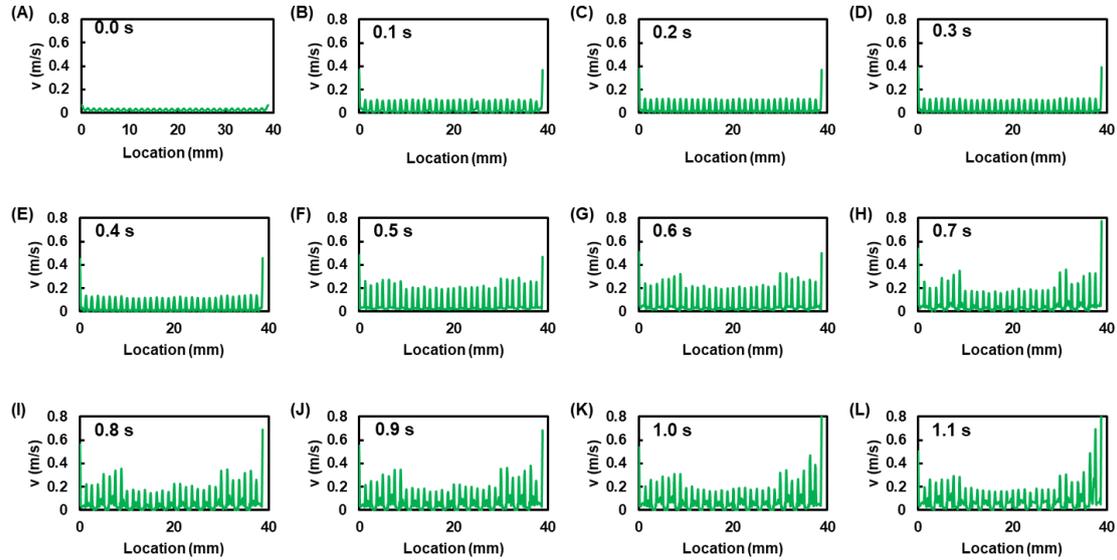

**Figure 4. The time evolution of fluid velocity across the outlet of the fractal structure at different time intervals from 0 to 1.1 seconds.** (A) The velocity profile at 0 seconds, where no significant flow is present. (B-F) the velocity begins to develop, with the fluid velocity gradually increasing across the outlet section. (F) The velocity exceeds 0.2 m/s. (G-L) The velocity stabilizes within a range of 0.2 m/s to 0.3 m/s, with slightly higher velocities observed near the edges of the outlet branches. The distribution of velocity becomes more prominent as time progresses, showing a generally uniform flow pattern across the fractal microchannels, with localized increases in velocity near the boundaries of the branches.

The results depicted in Fig. 4 illustrate the time evolution of the fluid's velocity distribution at the outlet of the fifth-level fractal structure. The observed gradual increase in velocity during the initial time frames (0 to 0.5 seconds) reflects the system's filling process, where fluid is entering the fractal network and begins flowing through the various microchannels. The relatively slow and uniform increase in velocity across the outlet at the early stages of the experiment indicates that the fluid experiences significant resistance within the fractal network, particularly within the smaller branches of the fractal structure. As time progresses and the flow reaches a steady state, the velocity distribution becomes more consistent, with higher velocities concentrated near the channel edges. This could be due to the geometrical constraints imposed by the fractal design, where the outermost branches of the structure allow for less resistance to fluid flow. These edge effects could be attributed to the boundary conditions imposed by the microchannel walls, where the interaction between the fluid and the surface increases shear forces, resulting in elevated velocity along the outer edges. The consistent velocity distribution between 0.5 seconds and 1.1 seconds suggests that the fractal structure effectively promotes even fluid distribution across the outlet, which is critical for applications requiring uniform flow, such as drug delivery and in vitro diagnostics. The design of the fractal network ensures that the fluid travels through multiple pathways, leading to a well-balanced flow profile at the outlet, which is essential for ensuring that treatments or diagnostic reagents are evenly distributed when exiting the system.

In this section, we analyze the time-dependent variation of the fluid mass fraction at the outlet cross-section of the fifth-level fractal structure. Figure 5 presents the mass fraction of the injected fluid at different time intervals, ranging from 0.0 s to 1.1 s, corresponding to the same outlet section discussed earlier in Figs. 3 and 4. At 0.0 s (Fig. 5a), the mass fraction of the injected fluid across the entire cross-sectional plane is zero, indicating that no fluid has yet passed through the outlet. This result is consistent with the velocity data shown in Fig. 4a, where the fluid velocity was near zero at this initial time point. Between 0.1 s (Fig. 5b) and 0.5 s (Fig. 5f), the mass fraction remains largely unchanged across the outlet. The negligible variation in fluid mass fraction during this time interval suggests that the injected fluid has not yet reached the outlet section. This observation correlates with the relatively slow fluid velocity in the fractal branches, as seen in the corresponding velocity profiles in Fig. 4b through 4f, where the fluid was only beginning to flow through the structure. The stagnation of the mass fraction during this period further supports the idea that the fractal network imposes significant resistance to the flow in its early stages. At 0.6 s (Fig. 5g), the mass fraction of the injected fluid begins to rise sharply, reaching approximately 0.2. This sudden increase marks the moment when the injected fluid reaches the outlet. The data suggests that the fluid first passes through the central branches of the fractal network before spreading to the outer branches, leading to the observed mass fraction rise. This moment corresponds with the notable increase in fluid velocity seen at 0.6 s (Fig. 4g), confirming that the flow has reached the outlet.

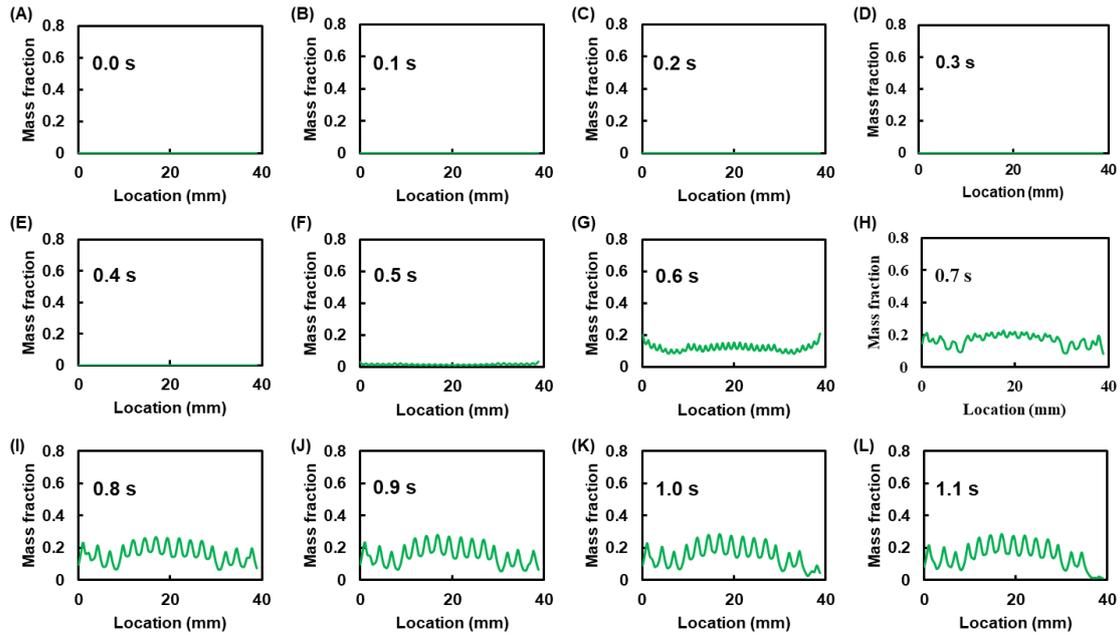

**Figure 5. Illustration of the time evolution of the mass fraction of the injected fluid at the outlet cross-section of the fifth-level fractal structure.** (A) At 0.0 s, the mass fraction is zero across the entire outlet, indicating that no fluid has reached the outlet at this time point. (B-F) From 0.1 s (b) to 0.5 s (F), the mass fraction remains unchanged, suggesting that the injected fluid has not yet arrived at the outlet. (G-L) At 0.6 s, the mass fraction increases to approximately 0.2, marking the point at which the fluid reaches the outlet. In the subsequent time intervals, from 0.7 s (H) to 1.1 s (L), the mass fraction stabilizes, with minor fluctuations reflecting the interaction between the fluid flow and the fractal network geometry. These fluctuations are more pronounced near the outlet edges, likely due to the reduced resistance in the peripheral microchannels, as indicated by the velocity data.

As time progresses, from 0.7 s (Fig. 5h) to 1.1 s (Fig. 5l), the mass fraction stabilizes, fluctuating around 0.2 across the outlet section. These minor fluctuations indicate localized variations in flow distribution due to the complex geometry of the fractal network, particularly at the boundaries of the microchannels. The overall mass fraction remains relatively uniform, with small oscillations reflecting the intricate interaction between the fractal geometry and the advancing fluid front. The higher fluid mass fraction near the outlet edges could be due to the peripheral microchannels exhibiting less resistance to flow, similar to the velocity distribution described in Fig. 4. The data from Fig. 5 highlight the influence of the fractal structure on fluid transport, with a delayed response in fluid mass fraction change due to the intricate path that the fluid must follow through the fractal branches. Once the fluid reaches the outlet, the mass fraction stabilizes, though slight fluctuations suggest ongoing interactions between the fluid and the complex geometrical features of the outlet. These findings underline the importance of fractal designs in achieving controlled fluid distribution, as the multi-branched structure ensures that the flow progresses uniformly through the system despite the complex network geometry.

Figures 6 and 7 provide a detailed analysis of the fluid mass fraction distribution across the outlet section of the fifth-level fractal structure. The study focuses on a subset of the total 32 outlets, specifically the left half of the structure, comprising 16 outlets symmetrically distributed, as depicted in Fig. 6. These outlets are highlighted in red, providing a clearer understanding of the regions analyzed. In Fig. 7, the fluid mass fraction across these 16 outlets is presented over time, covering the range from 0 to 1.2 seconds. The data is color-coded to reflect the mass fraction at different time points, with the corresponding outlets labeled from Fig. 7a (Outlet 1) to Fig. 7p (Outlet 16). The results indicate a relatively uniform distribution of the fluid mass fraction across the outlet section. In most cases, the curves show a nearly horizontal trend, suggesting a stable and consistent fluid distribution across the selected outlets. From the analysis, the fluid mass fraction stabilizes around 0.2, consistent with the data presented in Fig. 5, where the overall mass fraction also converged to approximately 0.2 at the outlet section. This uniformity in mass fraction distribution indicates that the fractal structure effectively promotes even fluid flow across multiple outlet branches, preventing significant variation in the mass flow rate between different outlets.

The small slope observed in certain curves may reflect minor flow discrepancies due to variations in microchannel geometry or slight differences in the resistance encountered by the fluid in the outermost branches of the fractal structure. However, these discrepancies are generally minor, as the mass fraction remains close to the average value of 0.2 across all outlets. The fractal design's ability to balance flow through such a complex network is crucial in applications where uniform fluid distribution is necessary, such as drug delivery systems or diagnostic assays. The uniformity in fluid mass fraction distribution, as observed in Fig. 7, further validates the efficacy of the fractal structure in managing the flow. As the fractal levels progress, the flow rate and fluid mass fraction are well-controlled across the smaller branches, ensuring that fluid delivery is consistent and predictable. This characteristic is particularly important in microfluidic applications where precise control over fluid distribution is critical for experimental outcomes, especially in sensitive processes such as immunoassays or drug delivery systems that depend on accurate dosing.

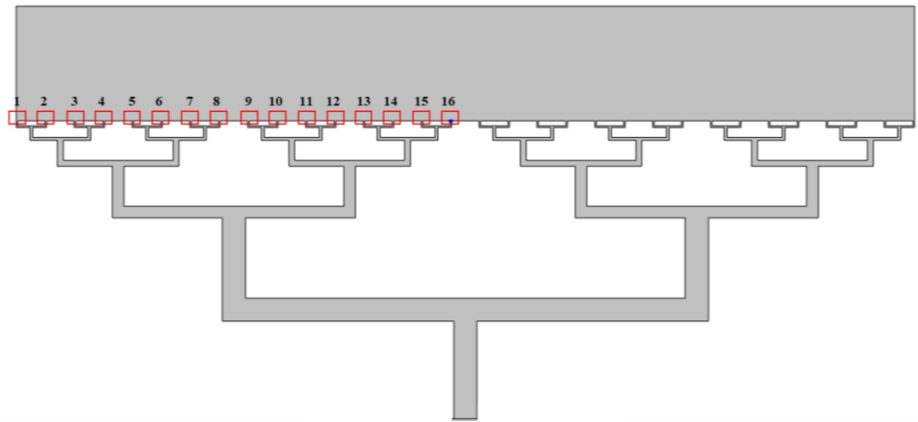

**Figure 6. The schematic of the fifth-level fractal structure, highlighting 16 outlets on the left side of the system, symmetrically distributed.** These outlets are selected for the analysis of fluid mass fraction, and are marked by a red box. The fractal structure at this stage consists of 32 outlets in total, with the 16 highlighted outlets representing half of the overall outlet branches.

The results from Figs. 6 and 7 highlight the effectiveness of the fractal design in ensuring a uniform fluid distribution across a multi-outlet structure. The slight variations in mass fraction between the outlets, while observable, are minimal and do not significantly affect the overall performance of the system. These minor deviations are likely caused by subtle differences in the microchannel geometry or small variations in the flow resistance encountered by the fluid in different outlets. The ability of fractal structure to achieve a uniform flow across a complex, multi-branch network is a significant advantage in applications that demand precise fluid control. For example, in immunoassays, maintaining a consistent flow rate and mass fraction is crucial for achieving accurate and reproducible results. Similarly, in drug delivery systems, ensuring that the drug is evenly distributed across all outlets ensures that each outlet receives an appropriate dosage, which is critical for the effectiveness of the treatment. The data also demonstrate that the mass fraction of the injected fluid stabilizes quickly after the initial flow reaches the outlets. The results presented in Figures 6 and 7 reinforce the utility of fractal networks in fluidic systems that require precise control over flow distribution. The fractal design not only ensures uniform mass fraction distribution but also facilitates rapid stabilization, making it an ideal candidate for applications in diagnostics, drug delivery, and other fields that rely on consistent and controlled fluid delivery.

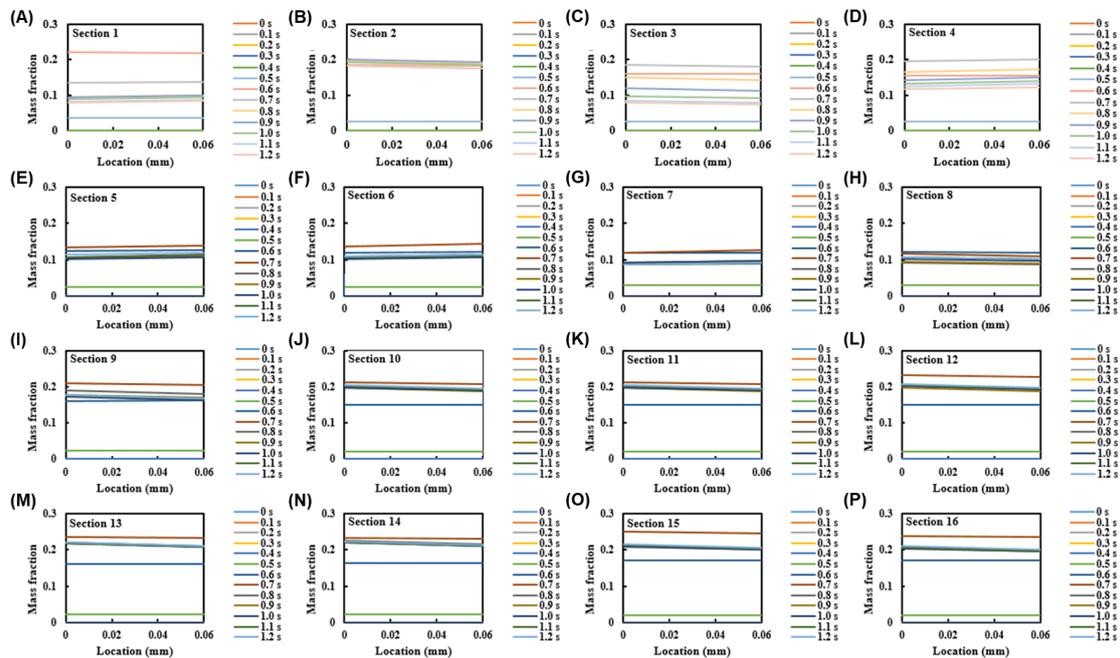

**Figure 7. The illustration of the time-dependent distribution of the fluid mass fraction at the 16 outlets highlighted in Figure 6.** (A-P) The data is shown from 0 to 1.2 seconds, with each sub-figure representing an individual outlet from Outlet 1 to Outlet 16. The fluid mass fraction is plotted across the outlet cross-section, and different colors are used to represent the mass fraction at different time points. The results indicate a generally

uniform mass fraction distribution across the outlets, with most curves maintaining a near-horizontal trend. The mass fraction stabilizes around 0.2, with minimal variations between outlets, suggesting consistent and even fluid flow across the fractal network.

**3.3. The uniformity of fluid flow at fractal outlet**

Figure 8 presents a detailed analysis of the influence of a flattened fractal structure on fluid distribution uniformity. The parameter RA, introduced in Fig. 8a, represents the ratio of the vertical pipe length to the horizontal pipe length at each level of the fractal structure. This ratio is essential in determining how fluid flows through the channels and the effect of the geometry on promoting even fluid distribution. The simulation models fluid flow in a third-level fractal structure with an inlet velocity of 0.5 m/s, and the temporal evolution of the flow is captured through finite element simulations. The velocity distribution is examined over time from 0 s to 0.5 s, offering a detailed view of how fluid flow stabilizes within the fractal structure. At 0 s (Fig. 8b), the initial velocity is close to zero across the structure, as expected without an applied flow. By 0.05 s (Fig. 8c), the velocity increases throughout the structure, with certain regions reaching velocities of around 0.05 m/s. By 0.1 s (Fig. 8d), the fluid velocity further increases, with values between 0.1 and 0.15 m/s spread evenly throughout the fractal channels. This shows a gradual acceleration of fluid flow across the structure.

By 0.15 s (Fig. 8e), the majority of the structure exhibits velocities of around 0.15 m/s, indicating that the fluid is continuing to accelerate and stabilize. At 0.2 s (Fig. 8f), the fluid velocity increases further, with some areas experiencing velocities exceeding 0.2 m/s, indicating the flow is approaching a stable state. By 0.25 s (Fig. 8g), the fluid flow achieves near-stability, with most regions displaying velocities around 0.25 m/s and some areas reaching speeds of 0.3 m/s. The simulation shows that the flow remains stable from 0.3 s to 0.5 s (Figs. 8h to 8l). During these periods, the velocity throughout the fractal structure remains constant, with maximum velocities ranging from 0.25 to 0.3 m/s. The stabilization of fluid velocity occurs within 0.25 s, demonstrating that the flattened fractal structure efficiently promotes uniform fluid distribution and rapid equilibrium. The results in Fig. 8 emphasize the role of the geometry of fractal structure in achieving uniform and stable fluid distribution. The parameter RA directly influences how quickly the fluid reaches a steady state, and in this case, the flow stabilizes within 0.25 s. This is a significant finding, as it underscores the importance of fractal geometry in controlling fluid dynamics. A particularly noteworthy aspect is the rapid stabilization of fluid velocity throughout the structure. This rapid equilibrium suggests that the flattened fractal design ensures efficient distribution of fluid, which is vital in applications requiring uniformity.

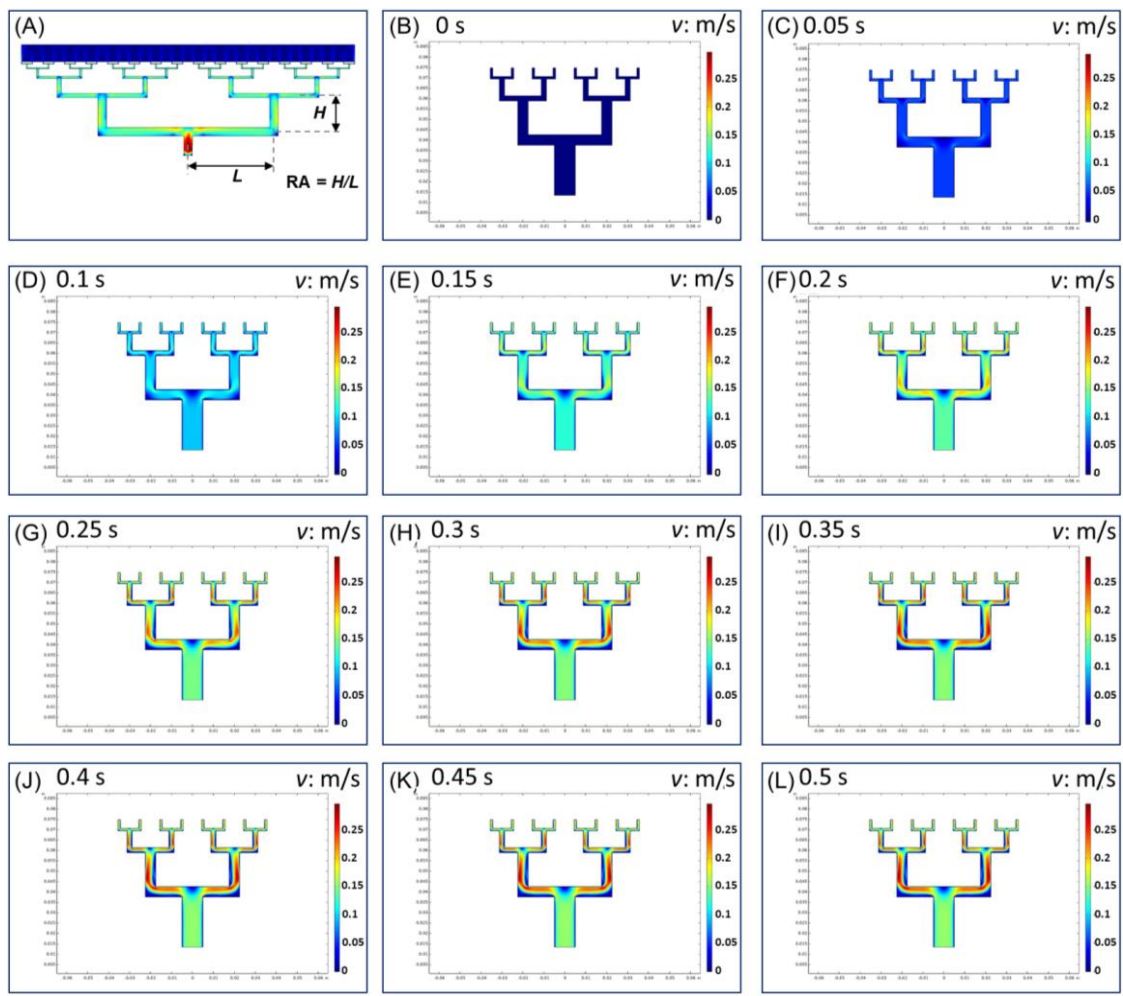

**Figure 8. The demonstration of the impact of the geometry of fractal structure on fluid velocity distribution.** (A) The parameter RA defined as the ratio of vertical to horizontal pipe lengths in the fractal structure. (B-L) Depiction of the finite element simulation results of fluid velocity distribution in a third-level fractal structure with an inlet velocity of 0.5 m/s. At 0 s (B), the initial velocity is near zero across the structure. By 0.1 s (D), the velocity increases to around 0.1 to 0.15 m/s. By 0.25 s (G), the velocity stabilizes, with most regions reaching speeds of 0.25

to 0.3 m/s. The flow remains stable from 0.3 s to 0.5 s (H-L), demonstrating the capability of fractal structure for uniform and rapid fluid distribution. The results indicate that the fluid flow stabilizes within 0.25 s.

Figure 9 presents a comparison of flow velocity uniformity at the outlets of a third-level fractal structure under different RA (the ratio of vertical to horizontal pipe lengths). The figure illustrates simulation results at steady-state, the coefficient of variation (CV) of velocity uniformity over time at the outlet, and the temporal evolution of flow velocity for each scenario. In Fig. 9a, where RA = 0.1, the maximum velocity after stabilization is around 0.17 m/s. The CV values remain low throughout the 2 s observation period, indicating excellent uniformity in fluid distribution across the outlets. This suggests that a smaller RA facilitates highly uniform flow, with minimal velocity variations at the outlet. In Fig. 9b, where RA = 0.2, the maximum velocity increases to around 0.23 m/s. The CV values again stay below 0.01, confirming that even at this higher velocity, the fluid distribution remains very uniform. The consistency in the low CV values indicates that the fractal structure is capable of maintaining flow uniformity despite the increase in velocity, as the influence of the higher RA value is not significant at this level. However, in Fig. 9c, where RA = 0.5, an obvious change in the uniformity is observed. The maximum velocity reaches 0.27 m/s, but the CV values exhibit more pronounced variation. The CV rises from 0 s and stabilizes around 0.02 at 0.5 s, suggesting that while the flow becomes stable, there is a slight decline in uniformity compared to the lower RA cases. The higher velocity at RA = 0.5 seems to introduce a degree of non-uniformity in the flow distribution at the outlet. In Fig. 9d, where RA = 1, the maximum velocity increases further to 0.3 m/s. The CV values show a noticeable rise from the start, stabilizing around 0.02 at approximately 0.7 s. This increase in CV indicates a further reduction in flow uniformity at the outlet compared to the lower RA values. The fluid velocity distribution becomes more uneven as the vertical-to-horizontal length ratio increases, suggesting that RA = 1 introduces a significant challenge to maintaining uniform flow distribution. Finally, in Fig. 9e, where RA = 2, the maximum velocity reaches 0.34 m/s. The CV stabilizes around 0.06 after approximately 0.7 s, indicating a further drop in the uniformity of fluid distribution. The high CV value demonstrates that as RA increases, the uniformity of fluid velocity across the outlet deteriorates. Although higher velocities are achieved, the trade-off is a loss in even flow distribution.

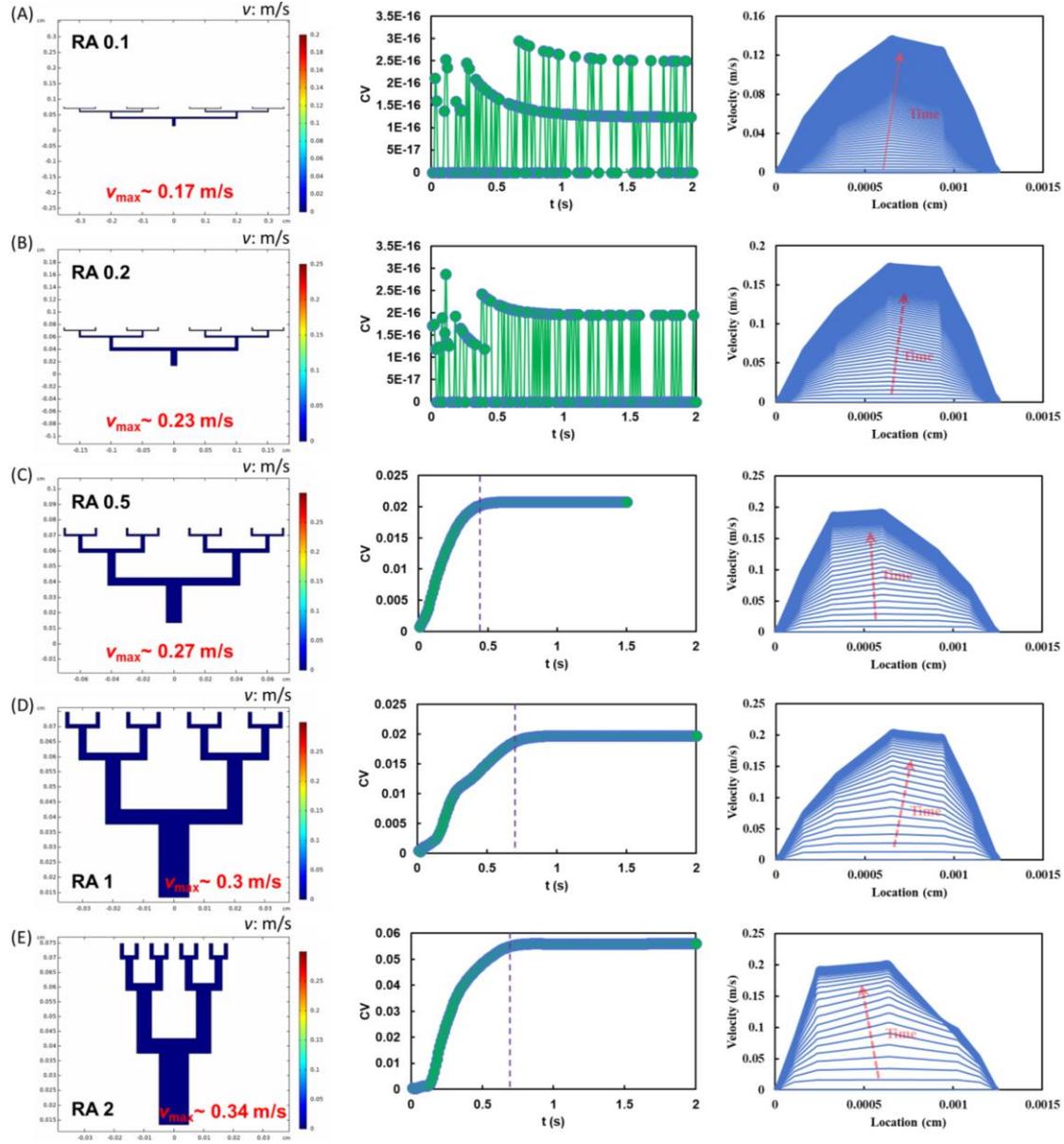

**Figure 9. The effect of different RA values on the flow velocity uniformity at the outlets of a third-level fractal structure.** Each subfigure presents the steady-state simulation results, the CV of velocity uniformity over time, and the temporal evolution of outlet flow velocity. (A) RA = 0.1, the maximum velocity is 0.17 m/s, with low CV values indicating excellent uniformity. (B) RA = 0.2, the maximum velocity is 0.23 m/s, with similarly low CV values. (C) RA = 0.5, the maximum velocity is 0.27 m/s, but the CV rises to 0.02, indicating reduced uniformity. (D) RA = 1,

the maximum velocity is 0.3 m/s, with CV stabilizing at 0.02. (E) RA = 2, the maximum velocity is 0.34 m/s, with CV reaching 0.06, demonstrating a further decrease in flow uniformity.

The results depicted in Fig. 9 demonstrate the critical impact of the RA parameter on the uniformity of flow velocity at the outlets of a fractal structure. Smaller RA values, such as 0.1 and 0.2, result in excellent flow uniformity, as evidenced by low CV values throughout the observation period. The minimal variations in outlet velocities highlight the ability of fractal structure to distribute fluid evenly when the vertical pipe length is much shorter than the horizontal length. This suggests that in applications requiring high flow uniformity, using smaller RA values is advantageous. As RA increases, there is a clear trend towards reduced flow uniformity. At RA = 0.5, while the flow still stabilizes, the CV values show increased variability, indicating that the flow velocity across the outlets becomes less uniform. This trend continues with RA = 1 and RA = 2, where the CV values rise significantly, reaching as high as 0.06. These findings suggest that as the vertical-to-horizontal pipe length ratio increases, the fractal structure struggles to maintain uniform flow, likely due to increased resistance and uneven distribution paths in the more vertical sections. The implications of these results are important for the design of fractal structures, especially in systems where uniform fluid distribution is critical. Applications such as microfluidic devices, chemical reactors, or cooling systems that rely on consistent flow at multiple outlets would benefit from lower RA values to ensure even distribution. However, if higher velocities are desired, a balance must be struck between velocity and uniformity, as increasing RA enhances flow speed but compromises outlet uniformity.

## 4. Conclusion

In conclusion, the fluid flow dynamics and distribution uniformity in dendritic fractal structures were studied using finite element analysis. The investigation focused on velocity distribution, mass fraction changes, and flow patterns at the outlets of fractal channels with varying structural proportions. Results indicated that the flow velocity and the mass fraction in the outlets are relatively uniform in the cross-sections of outlets.

The effect of different RA values, the ratio of vertical to horizontal pipe lengths, on outlet velocity uniformity were explored. Lower RA values (e.g., 0.1 and 0.2) resulted in highly uniform velocity distributions, with low CV values indicating minimal variation across the outlets. As RA increased to 0.5 and beyond, velocity distribution became less uniform, and CV values showed noticeable increases, demonstrating a trade-off between flow velocity and uniformity. The results also showed that fluid velocity stabilized quickly, typically within 0.25 s for smaller RA values, which suggests efficient fluid transport. However, the loss of uniformity at higher RA values indicates that lower RA designs are better suited for applications requiring consistent fluid dispersion. In summary, the study highlights the critical role of the RA parameter in controlling the uniformity and velocity of fluid flow in fractal structures. Smaller RA values lead to more uniform distribution, while larger RA values promote higher velocity but reduce uniformity. These insights provide useful guidelines for designing fractal structures in engineering and biomedical fields where precise fluid control is needed.

## Acknowledgment

This work was supported by the National Natural Science Foundation of China (No. 12202461).

## Conflict of Interest

The authors declare no competing financial interest.